\documentclass{iopart}

\newcommand{\wvec}[1]{\protect{\overline{#1}}}
\usepackage{iopams}
\usepackage{graphicx}
\begin{document}
\title{Aspects of Causality in Parallelisable Implicit Evolution Scheme}

\author{P. Khavari$^1$, C. C. Dyer$^{1,2}$}
\address{$^1$ Department of Astronomy and Astrophyscis, University of
Toronto, 50 Saint George Street,\\Toronto, ON, M5S 3H4, Canada}
\address{$^2$ Department of Physical and Environmental Sciences,
University of Toronto at Scarborough, 1265 Military Trail, Toronto, ON,
M1C 1A4, Canada}
\ead{khavari@astro.utoronto.ca}
\ead{dyer@astro.utoronto.ca}

\begin{abstract}
A (3+1)-evolutionary method in the framework of Regge Calculus,
essentially a method of approximating manifolds with rigid simplices,
makes an excellent tool to probe the evolution of manifolds with
non-trivial topology or devoid of symmetry. The
``Parallelisable Implicit Evolution Scheme'' is
one such method. Causality however, is an aspect of this method that has
been barely investigated. In this paper, we show how causality can be
accounted for in this evolutionary scheme. The revised algorithm is
illustrated by a preliminary application to a skeletonised spherical
Friedmann-Lema\^itre-Robertson-Walker universe.
\end{abstract}

\submitto{\CQG}



\section{Introduction}
Regge Calculus \cite{Ref1} is a numerical method in general relativity
that trades curved manifolds for skeletonised space-times. Considered
a finite element method, Regge Calculus approximates a manifold using
rigid simplices, higher dimensional generalisation of triangles
and tetrahedra. Once an n-dimensional manifold is approximated by
n-dimensional simplices, the intrinsic curvature is concentrated on
sub-simplices of co-dimension two; the geometry is then flat everywhere
within the simplices. The amount of curvature residing on the so-called
hinge or bone (sub-simplex of co-dimension two) is then represented by the
``deficit angle'' associated with the bone \cite{Ref1}. In this formalism,
the dynamical variables
are the edge lengths of the simplicies which indeed play the role of
the metric in the continuum limit. \\

In addition to being a powerful tool for investigation of problems devoid
of symmetry, today Regge Calculus is considered a promising approach in
the search for a theory of Quantum Gravity (see the extensive review by
Hamber \cite{Ref2}). We believe that Regge Calculus is also an excellent
method to incorporate non-trivial topology into general relativity, which
only provides one with local geometry and gives very little information
about global features of a manifold including its topology.\\

One can examine the evolution of a space-time with non-trivial topology
using a (3+1)-evolutionary method in the context of Regge Calculus. One
such evolutionary scheme, entitled ``Parallelisable Implicit Evolution
Scheme'', was presented by Barrett et al. \cite{Ref3} based on an
earlier work by Sorkin \cite{Ref4}. Causality is an aspect of this
algorithm that has not been investigated thoroughly and it has not been
incorporated into this algorithm properly. In their seminal paper on
``Parallelisable Implicit Evolution Scheme'' (PIES), also known as ``Sorkin
Triangulation'', considering the particular method of triangulation and
the causal structure of the skeletonised manifold, Barrett et al. place
a restriction on the type of edges that connect two consecutive
triangulated spatial hypersurfaces so that causality is not violated.
They indicate, however, that causality, in their scheme, is an issue
that requires further investigation. As will be described later, the restriction imposed by Barrett
et al. indeed prevents the time-like paths of evolving vertices from
colliding with their neighbors, but this restriction does not account
for causality. PIES, as presented by Barrett et al., does not produce the expected results when implemented to examine the evolution of a skeletonised Friedman-Lema\^itre-Robertson-Walker spherical universe. In particular, the evolution of this lattice universe stops well before the spatial volume becomes zero. This problem is known as the ``stop point" problem. We are confident that the ``stop point" problem arises because causality is not accounted for in PIES.\\

In this paper, we show how causality can be incorporated into PIES
properly and what changes must be made so that the Courant condition is
satisfied. The inclusion of causality, as done in this paper, resolves the ``stop point" problem as
will be described in detail in section (\ref{Area of a Bone
and the Issue of Causality}). We believe
that by accounting for causality, we have brought the ``Parallelisable
Implicit Evolution Scheme'' a step closer into being a highly efficient
tool in examining the evolution of skeletonised manifolds.\\

Section (\ref{A Brief Review of ``Parallelisable Implicit Evolution
Scheme'' Approach}) briefly reviews the ``Parallelisable Implicit
Evolution Scheme'' and describes the problem of causality. Sections
(\ref{Area of a Bone and the Issue of Causality}) and (\ref{Variation of a Time-Like Bone with respect to a Time-Like edge}) discuss how
causality can be included in this evolutionary scheme. In section (\ref
{A Simple Numerical Example}), we illustrate the revised algorithm by
applying it to successfully reconstruct a skeletonised spherical
Friedmann-Lema\^itre-Robertson-Walker (FLRW) universe. Section
(\ref{Conclusion}) closes this paper with some concluding remarks.


\section{A Brief Review of ``Parallelisable Implicit Evolution Scheme''
Approach}\label{A Brief Review of ``Parallelisable Implicit Evolution
Scheme'' Approach} As described in the introduction, Barrett et al. developed a (3+1)-evolutionary method in the context of
Regge Calculus \cite{Ref3}.
The so-called ``Parallelisable Implicit Evolution Scheme'',
considers a triangulated spatial section and assumes that all the
information pertaining to the triangulation of the space-time up to and
including this particular spatial hypersurface is known. It then
prescribes how one can evolve this skeletonised spatial hypersurface
into the future.\\

\begin{figure}[htbp]
\begin{center}$
\begin{array}{cc}
\includegraphics[scale=0.4]{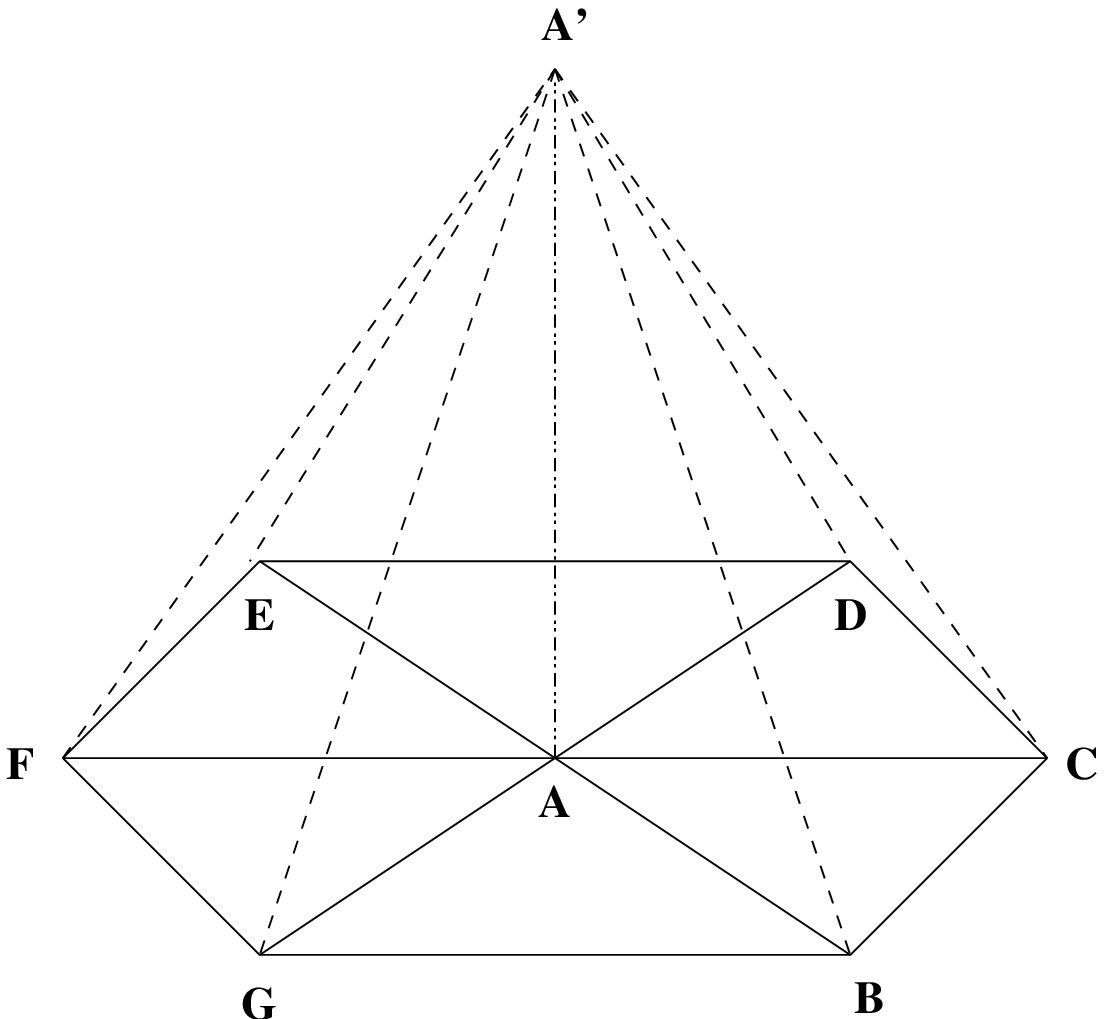} &
\hspace{0.5in}
\includegraphics[scale=0.4]{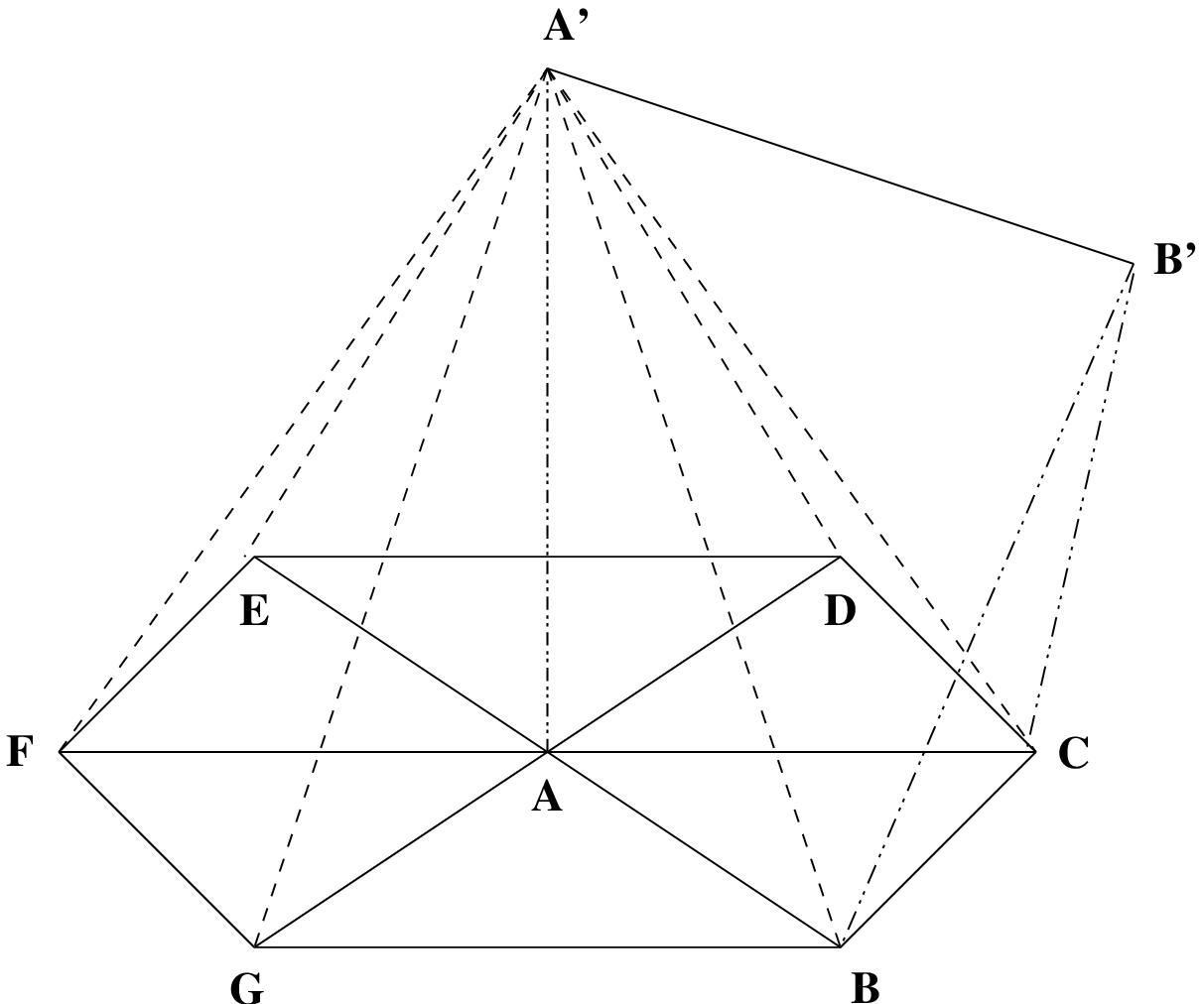}
\end{array}$
\end{center}
\caption{\label{Algorithm Illustration}
{\bf{Left:}} $A'$ is chosen to be the evolved counter-part of
$A$. $A$ and $A'$ are connected via a time-like ``vertical" edge. The rest
of the vertices, i.e. $B$, $C$, ... are connected to $A'$ via space-like
``diagonal" edges. Bones with two space-like and one time-like sides
(SST) are formed in all triangles with one vertical and one diagonal
edge. The length of 4 of the edges can be chosen arbitrarily.\\
{\bf{Right:}} $B'$ is the evolved counter-part of $B$. Again, $BB'$ is
a time-like edge but all other vertices, including $A'$ are connected
to $B$ by space-like edges. $A'B'$ is the evolved counter-part of edge $AB$.}
\end{figure}
More specifically, this method starts with an arbitrary vertex on
the initial hypersurface.  It then introduces an evolved counter-part
to this vertex in the temporal direction.  In figure (\ref{Algorithm
Illustration}), suppose vertex $A'$ is the evolved counter-part of vertex
$A$ (the un-primed vertices all reside on a spatial skeletonised foliation
of a space-time). According to this algorithm, $A$ and $A'$ are to be
connected by a ``vertical'' edge. In addition, all the vertices that were directly connected to $A$ on the initial hypersurface, are to be connected to $A'$ by ``diagonal'' edges.  Thus, if $A$ were directly connected to $n$
other vertices on the initial hypersurface, one step of this algorithm
produces $(n+1)$ struts that go between this hypersurface and the one
to be built ``above'' it. The length of four of these newly added edges
can be chosen arbitrarily, corresponding to the freedom in the choice of
lapse and shift \footnote{The ADM formalism, developed by Arnowitt, Deser and Misner, foliates space-time into space-like hypersurfaces. The lapse function and shift vector indicate how these foliations are welded to each other. The freedom in choice of lapse and shift arises from the Bianchi identities. It is well known that Bianchi identities have a counterpart in Regge Calculus and this provides one with the advantage of choosing four of the edge lengths arbitrarily, similar to what is done in ADM formalism.}. Solving the relevant Regge equations, one obtains the
length of the rest of the struts. To build and later find the length of the edges in
the next spatial hypersurface, one has to continue the above-mentioned
procedure for each of the vertices of the initial hypersurface. The next
step is very similar to this step, differing from it only in that the two new vertices, i.e. $A'$ and $B'$, must also be connected. As shown in figure (\ref{Algorithm
Illustration}), $A'B'$ is the first edge of the next spatial hypersurface;
the evolved counterpart of $AB$.\\

To our knowledge, PIES is by far the most efficient (3+1)-evolutionary
method introduced in the context of Regge Calculus. The evolution of a space-like hypersurface can be achieved locally by evolving one vertex at a time and in parallel for those vertices which are not directly connected. 
\begin{figure}[htbp]
\begin{center}
\includegraphics[scale=0.4]{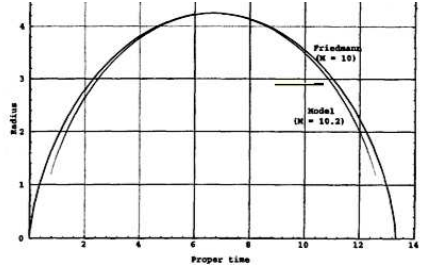} 
\caption{\label{Barrett's Figure}The Evolution of the spherical FLRW Universe using
the 600-cell triangulation of 3-sphere as obtained by Barrett et al. The
evolution stops well before reaching zero spatial volume \cite{Ref3}.}
\end{center}
\end{figure}
Other evolutionary
methods developed based on Regge Calculus use non-simplicial blocks.
This is indeed a disadvantage as these methods require the introduction
of information other than the edge lengths of the building blocks. PIES,
however, has a big shortcoming: when Barrett et al. try to use this
method to examine the evolution of a skeletonised spherical
Friedmann-Lema\^itre-Robertson-Walker Universe, they find that the
evolution stops well before the spatial volume of this universe even
gets close to zero. A graph of the evolution of the skeletonised FLRW, as obtained by Barrett at al., is shown in figure (\ref{Barrett's Figure}). We believe that this problem has root in not accounting for causality correctly. In this paper, we show that by correctly accounting for causality the ``stop point" problem will be resolved.\\

\section{Area of a Bone and the Issue of Causality}
\begin{figure}[htbp]
\begin{center}
\includegraphics[scale=0.4]{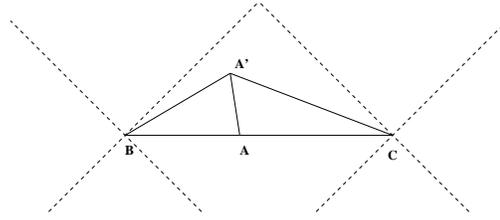}
\caption{\label{NoCollision} An illustration of PIES in a (1+1) skeletonised space-time.
Barrett et al. require that the diagonal edges such as $BA'$
be space-like while the evolutionary paths of vertices, such as $AA'$
can be time-like, space-like or null. This condition only prevents the
time-like evolutionary paths of vertices not to collide. This condition
is a ``No Collision'' condition which results in a piece-wise linear
congruence of non-intersecting paths of evolving vertices.}
\end{center}
\end{figure}

\label{Area of a Bone
and the Issue of Causality} In their seminal paper on PIES, Barrett et
al. argue that since in this approach one tries to obtain the
information about the newly introduced edges from the knowledge of the
triangulation of the initial spatial hypersurface, the tent-like
structure formed above a chosen vertex on a spatial hypersurface must
reside within the future domain of dependence of this hypersurface.
Thus, as shown in figure (\ref{NoCollision}), the diagonal edges must be
space-like while the vertical edge can in principle be time-like, null
or space-like. The restriction as imposed by Barrett et al. is more a
``No Collision'' requirement than a causality requirement. What this
condition does is that by making the diagonal edges space-like, prevent the time-like paths of evolving vertices from colliding in the
future.\\

\begin{figure}[htbp]
\begin{center}
\includegraphics[scale=0.4]{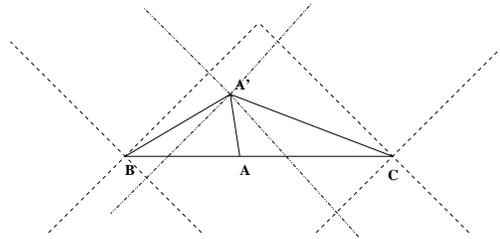}
\caption{\label{Causality} Only the information within the past null cone of $A'$ could have affected it.}
\end{center}
\end{figure}

\begin{figure}[htbp]
\begin{center}
\includegraphics[scale=0.4]{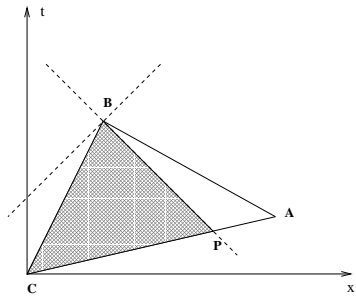}
\caption{\label{triangle_1.eps}
Vertex $B$ is taken to be the evolved version of vertex $C$. Only
the crossed-hatched area of triangle $\triangle{CAB}$ is within the null
cone of vertex $B$.}
\end{center}
\end{figure}

To include causality without violating the above-mentioned
condition, one has to look at the past null cone of the evolved
counterpart of a vertex.
 Figure (\ref{Causality}) shows the situation in a (1+1)-dimensional space-time. It is quite clear that not the entire 1-d piece-wise linear space is within the past null cone of vertex $A'$.
It is best to discuss causality in (3+1)-dimensional skeletonised space-times.
Consider triangle $\triangle{CAB}$ in figure (\ref{triangle_1.eps});
suppose that $CA$ is a space-like edge on a triangulated 3-dimensional
spatial hypersurface. Following the PIES algorithm, assume vertex $B$
is the
evolved counterpart of vertex $C$. Edge $BC$ is time-like but edges $AB$ and $CA$ are space-like as prescribed by the algorithm. 
$CA$ resides on the initial hypersurface while $BC$ and $AB$ go between the two
hypersurfaces. 
The null cone of vertex $B$ divides the time-like
bone \footnote{A space-like bone is a bone made of only space-like
edges. A time-like bone however is constructed from a combination of
both time-like and space-like edges.}, $\triangle{CAB}$, into a triangle
with two space-like and one null edge (NSS) and a triangle with one
time-like, one space-like and one null edge (NST). Clearly, only the
(NST) part of the bone is in the past domain of dependence of vertex $B$
and could have had any influence on $B$. Thus to account for causality,
we have to include this fact in the action.\\

It is well known that the action for a skeletonised space-time is given
by \cite{Ref1}:
\begin{equation}
I = \sum_{n}{A_{n} \epsilon_{n}}
\end{equation}
The Regge equation is obtained by varying this action with respect to a
given edge. In his seminal paper, Regge showed that one can carry out
this variation as if the deficiencies were constant.
The skeletonised version of Einstein's equation is then given by:
\begin{equation}
\sum_{n}{\frac{\partial{A_{n}}}{\partial{L_{i}}} \epsilon_{n}} = 0
\end{equation}
To include causality in the ``Parallelisable Implicit Evolution Scheme'',
instead of the entire area of the bone in Regge action, only
the part which is within the past null cone of vertex $B$ must be included in the action. In particular,
in writing the relevant Regge equations obtained by varying the area
of a bone with respect to $CB$, one has to carry out this variation for
the area of the (NST) triangle, $\triangle{CBP}$.
We now carry out this variation for a time-like bone. A similar line of
argument can be used to obtain similar results for a space-like bone.

\section{Variation of a Time-Like Bone with respect to a Time-Like
edge}\label{Variation of a Time-Like Bone with respect to a Time-Like
edge}
\begin{figure}[htbp]
\begin{center}
\includegraphics[scale=0.4]{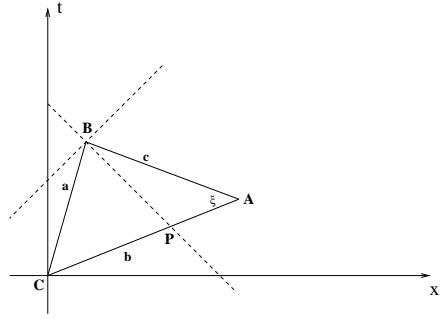}
\end{center}
\vspace*{8pt}
\caption{\label{Triangle4}
The time-like bone $\triangle{CAB}$ is divided into a NST and
a NSS triangle by the null line passing through $B$.}
\end{figure}

Consider the time-like bone $\triangle{CAB}$ in figure
(\ref{Triangle4}). In this triangle we have:
\begin{displaymath}
A_{\triangle{CBP}} = A_{\triangle{CAB}} - A_{\triangle{APB}}
\end{displaymath}
From (\protect \ref{Area CBP}), one has:
\begin{equation}
\label{Area of CBP}
A_{\triangle{CBP}} =  A_{\triangle{CAB}} -
\frac{A_{\triangle{CAB}}}{2b^2}(a^2+b^2+c^2) +
\frac{4A^2_{\triangle{CAB}}}{2b^2}
\end{equation}
Varying the area of $\triangle{CBP}$ with respect to ``$a$", the time-like
edge of $\triangle{CAB}$, one has:
\begin{equation}
\label{variation1}
\frac{\partial{A_{\triangle{CBP}}}}{\partial{a}} =
\frac{\partial{A_{\triangle{CAB}}}}{\partial {a}}(1-
\frac{a^2+b^2+c^2}{2b^2}+\frac{4A_{\triangle{CAB}}}{b^2})-
\frac{a}{b^2}A_{\triangle{CAB}}
\end{equation}
but
\begin{displaymath}
\frac{\partial{A_{\triangle{CAB}}}}{\partial{a}} =
\frac{1}{2}a\:\frac{(b^2 + a^2 + c^2)}{4{A_{\triangle{CAB}}}} =
\frac{1}{2}a\:\coth{\xi}
\end{displaymath}
Inserting this into equation (\protect \ref{variation1}) results in:
\begin{displaymath}
\frac{\partial{A_{\triangle{CBP}}}}{\partial{a}} =
\frac{1}{2}a\;(\coth{\xi})(1 -\frac{c}{b}\cosh{\xi} +
\frac{2c}{b}\sinh{\xi}) - \frac{a}{b^2}A_{\triangle{CAB}}
\end{displaymath}
where we have used
\begin{displaymath}
{4A_{\triangle{CAB}}}/{b^2} = ({2c}/{b})\sinh{\xi}
\;\;\;\;\;\;\; {\rm and} \;\;\;\;\;\;
(c/b)\cosh{\xi} = {(b^2 + a^2 + c^2)}/{2b^2}.
\end{displaymath}
One can simplify this equation by replacing $A_{\triangle{CAB}}$ with
$\frac{1}{2}b\:c\sinh{\xi}$ to obtain:
\begin{equation}\label{partial derivative of NST with respect to edge}
\frac{\partial{A_{\triangle{CBP}}}}{\partial{a}} =
\frac{1}{2}a\:(\coth{\xi} - \frac{c}{b}\: e^{-2\xi}\rm csch\;{\xi})
\end{equation}
Finally, generalising equation (\protect
\ref{partial derivative of NST with respect to edge}) for all the bones
hanging at edge $a$, one obtains the relevant Regge equation that must be used in a ``causal PIES":
\begin{equation}
\label{Regge's equation revised}
\sum_n \frac{1}{2}a\big[\coth{\xi_n} - \frac{c_n}{b_n}\: e^{-2\xi_n}\rm
csch\;{\xi_n}\big] \epsilon_n = 0
\end{equation}
where the sum is over all the bones meeting at the time-like edge ``$a$"
and $\xi_{n}$ is the angle opposite to ``$a$" in the $n^{th}$ bone hanging at
edge ``$a$". $\epsilon_n$ stands for the deficiency associated with bone $n$.
\section{Numerical Examples}\label {A Simple Numerical Example}
In this section, we illustrate the revised algorithm by examining two
skeletonised spherical FLRW universes. These numerical example are very
close in nature to that given by Barret et al. and the details are quite
similar. The interested reader is referred to section (6) of Barrett et al. paper \cite{Ref3}. The major difference between our solution and that of Barrett et al. is that, in obtaining these solutions, we have used our revised equation
as obtained in the previous section.
\begin{figure}[htbp]
\begin{center}$
\begin{array}{cc}
\includegraphics[scale=0.4]{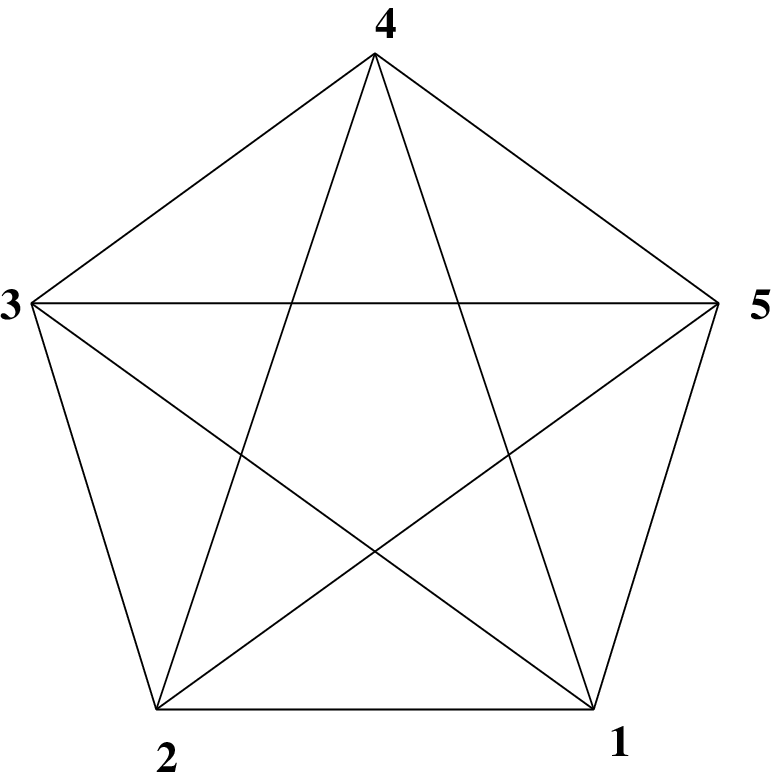}&\hspace{1.5cm}\includegraphics[scale=0.4]{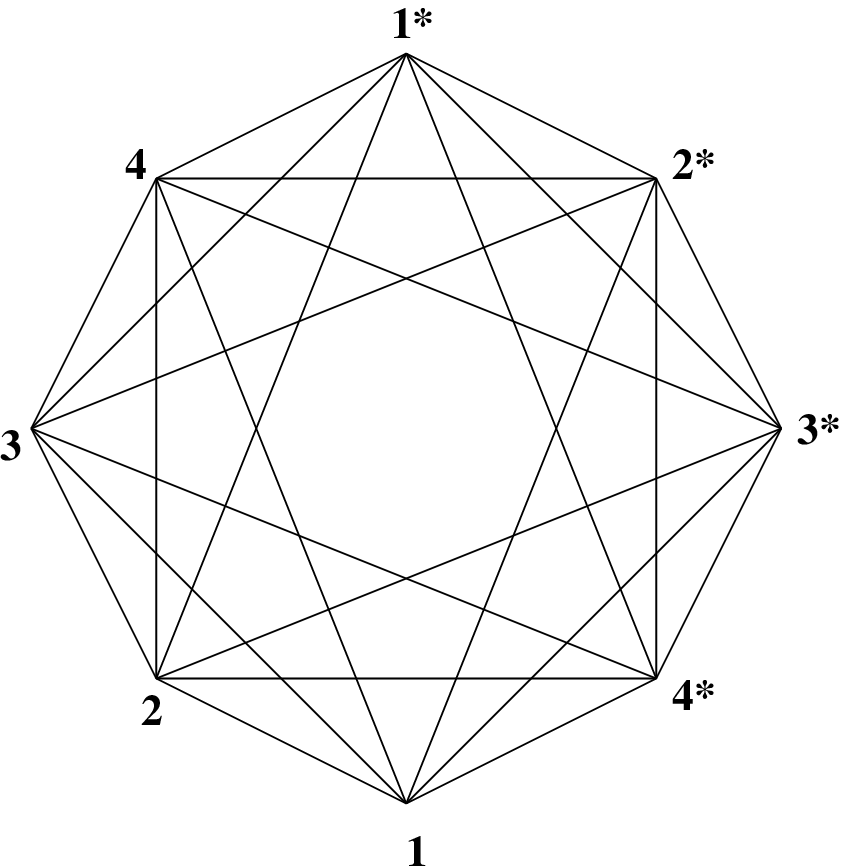}
\end{array}$
\caption{\label{Pentatope}
Pentatope (left) and Hexadecachoron (right) are standard
triangulations of a 3-sphere.}
\end{center}
\end{figure}
We have chosen the surfaces of a pentatope (5-cell) as well as a
hexadecachoron (16-cell), which are simple standard triangulations
of a 3-sphere, shown in figure (\ref{Pentatope}),
as our underlying lattices. In addition, our choice of time function,
as will be described below, is different from that of Barrett et al.\\

The evolution of the two models is very similar and thus we choose to
discuss the pentatope universe.
Assembling five dust particles on each of the vertices of a pentatope,
we use the revised algorithm to evolve a given hypersurface in time. Each
dust particle is taken to have a mass of $M/5$ where $M$ is the total
mass of the skeletonised universe. To compare the evolution of this
skeletonised universe with the analytical solution, an ``effective radius"
is introduced. This effective radius or scale factor of the lattice universe
is obtained by equating the volume of the pentatope to the analytical
volume of a 3-sphere at each step. More specifically, we have:
\begin{equation}
\frac{5l^{3}}{6\sqrt{2}} = 2\pi^{2}a_{e}^{3}
\end{equation}
where $l$ is the triangulation edge length and $a_{e}$ is the so-called
effective radius. We follow the contraction of this lattice universe
from the moment of time symmetry on \cite{Ref5}.\\

Another important issue in comparing the evolution of the skeletonised
universe with the analytical solution is the choice of a correct time
function. A straightforward analysis of the Robertson-Walker metric shows
that the elapsed time between two consecutive skeletonised spatial hypersurfaces in continuum
is given by the change in the 4-volume divided by the 3-volume of the
initial hypersurface. Extending this analysis to the discretised regime,
the lapse of time must be given by the 4-volume of the object sandwiched
between two consecutive hypersurfaces, divided by the 3-volume of the
base. However, the
nature of the algorithm is such that this block has a very complicated
construction and calculating the 4-volume is extremely hard. As a matter
of fact, as will be discussed below, the notion of proper time is much
more complicated in skeletonised space-times. \\

\begin{figure}[htbp]
\begin{center}
\includegraphics[scale=0.5]{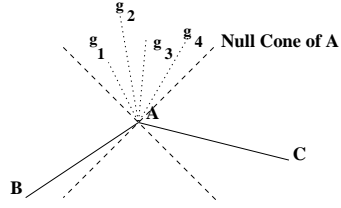}
\caption{\label{Arc2}
Any line segment emanating from vertex $A$ and lying within the future
null cone of $A$ is a portion of a future-pointing time-like
geodesic. The length of each of these is the proper time measured by the
observer moving along that time-like line, but none can be preferred over
the others.}
\end{center}
\end{figure}

At first glance, the norm of the time-like edge [16] might appear to be a
good candidate to represent the lapse of time. However this choice fails
because of the very nature of the algorithm: as the curvature in a
skeletonised space-time is concentrated on hinges, the space-time is
flat everywhere else and thus all the line segments emanating from a
vertex are portions of geodesics as shown in figure (\ref{Arc2}). The
length of each of these line segments correspond to proper time
intervals measured by different observers. These times are related to
one another by Lorentz transformations. It is not however possible to
distinguish which one of these observers is measuring the comoving time,
since for each such observer, their proper time defines what they mean
by cosmic time. Since there is not a unique
comoving observer at each vertex, but a class of such observers, there
is no unique comoving time, but a whole class of choices of comoving
time. The notion of orthogonality at a vertex, in a
piece-wise linear regime, is not well defined either. More specifically, the notion of an orthogonal vector at a cone singularity is not well defined. Thus it is not feasible to define a unique
comoving time as is done in the continuous case using the Weyl
postulate.  Consequently, it is not possible to define a unique notion
of comoving time for skeletonised space-times as in the continuum.\\

\begin{figure}[htbp]
\begin{center}$
\begin{array}{cc}
\includegraphics[scale=0.4]{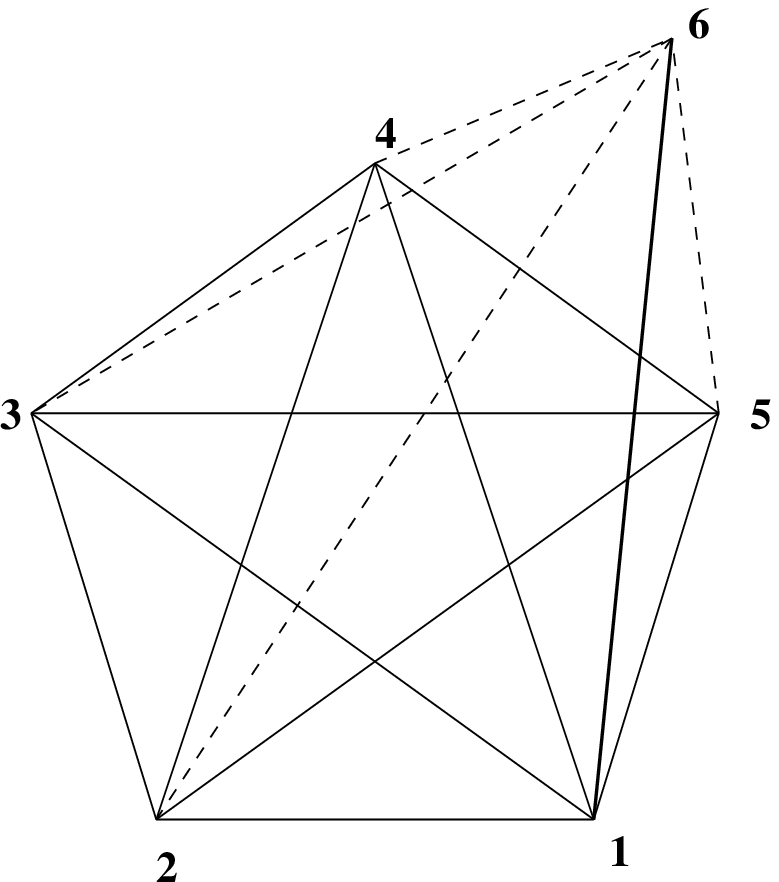} &
\hspace{1in}
\includegraphics[scale=0.4]{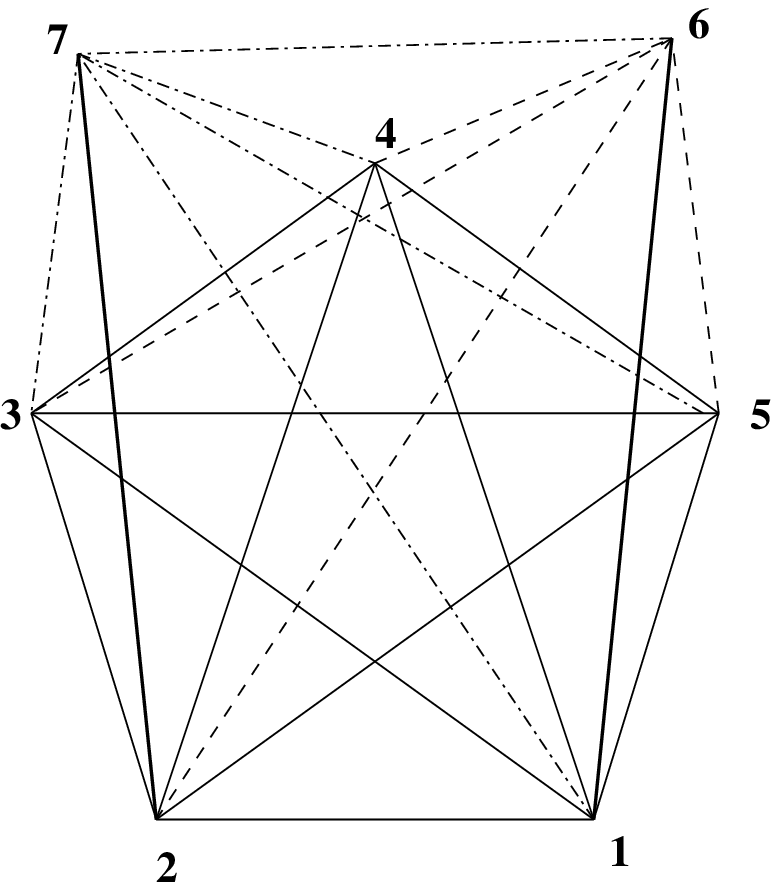}
\end{array}$
\caption{\label{EvolveVertex}Evolving vertex [1] to [6] produces five 4-simplices, four of
them contain the time-like edge [16] and are all equivalent. The one that does not contain edge [16]
is purely space-like. The elapsed proper time is taken to be proportional
to the 4-volume of one of the 4-simplices with a time-like edge.}
\end{center}
\end{figure}

One solution to this problem is that instead of using the exact lapse of time, a measure of this
lapse be obtained from the properties of the lattice that change with the
evolution. We choose to represent the lapse of time with the volume
of part of the 4-dimensional block, scaled by the edge length of the
triangulation of the hypersurface that is being evolved. The volume of a
4-dimensional simplex, such as [12346], as illustrated in
figure(\ref{EvolveVertex}), is ideal. This particular choice of
time, contains many properties of the initial hypersurface and its evolved
counter-part ([1234] is a tetrahedral block in the initial hypersurface
and [16] is an evolutionary step which is time-like). It is well known
that the 4-volume is indeed a reasonable choice to represent the lapse of
time \cite{Ref7}.  We believe that our choice of time is appropriate as it
embraces many evolutionary features of the algorithm. This choice however
is proportional to the proper time and the constant of proportionality
is a free parameter in our model. \\

To compare the revised algorithm with PIES, we make similar assumptions
to Barrett et al. in that we take all the diagonal edges to be of
equal length. In addition, we take the pentatopes, corresponding to the
triangulation of a spatial 3-surfaces to be equilateral. Finally, all the
vertical edges, connecting a vertex to its evolved counterpart, are taken to have equal lengths. These assumptions indeed
correspond to a skeletonised isotropic and homogeneous universe. Following
the algorithm and using the revised equations, one obtains two
roots for the length of diagonal edges, one that corresponds to a
contracting universe and one corresponding to a universe which expands
indefinitely. Choosing the former and applying the algorithm one more
time, say to vertex [2], we obtain two roots for the triangulation edge
length of the next spatial hypersurface. The difference between the
two roots, obtained in this step, is of the order of $10^{-2}$ and both lead into acceptable
solutions.\\

\begin{figure}[htbp]
\begin{center}$
\begin{array}{cc}
\includegraphics[scale=0.6]{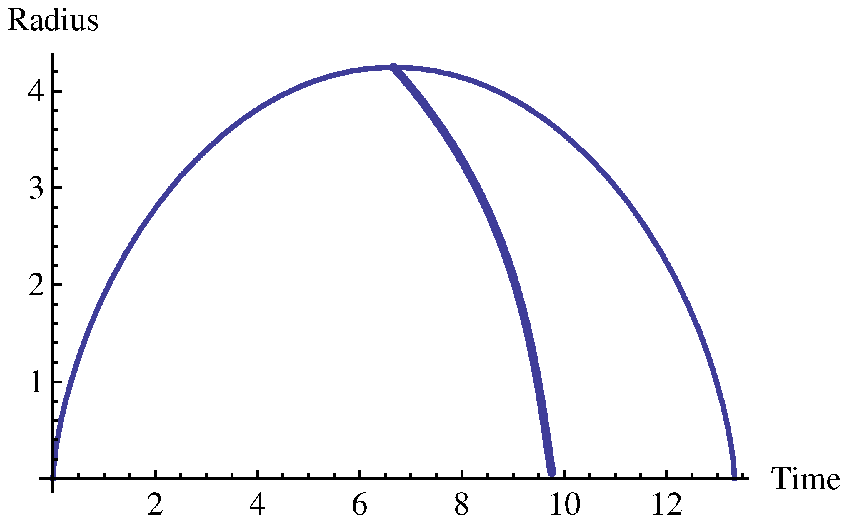} &
\includegraphics[scale=0.6]{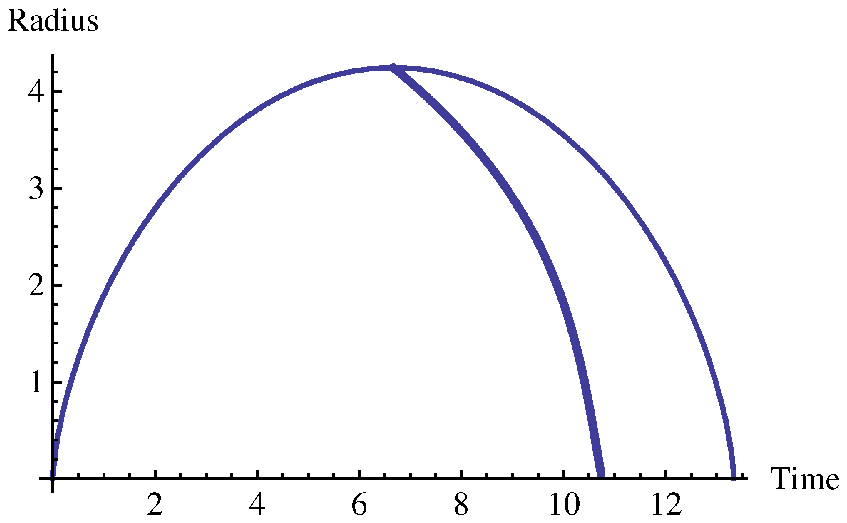}
\end{array}$
\caption{\label{5and16-CellGraphLR}
Left Panel: The Evolution of the Pentatope FLRW Universe. The mass
of the analytical solution is taken to be M = 10. The mass of the
skeletonised universe is 14.001 and the evolution steps are equal to
0.01. The constant of proportionality in the time-function is taken to
be 1/20.\\
Right Panel: The Evolution of the Hexadecachoron FLRW Universe. The
mass of the analytical solution is taken to be M = 10. The mass of the
skeletonised universe is 12.052 and the evolution steps are equal to
0.01. The constant of proportionality in the time-function is taken to
be 1/4.}
\end{center}
\end{figure}

The most important feature of this solution is that, independent of the choice of time function, the evolution does not stop at a finite
spatial volume. This is indeed firm evidence for the fact that the
introduction of causality in the algorithm resolves the problem of stop
point. A curious feature of this solution is that the evolution becomes
slower and slower as the spatial volume gets smaller and closer to
zero. In particular, by taking the same evolutionary step, the change
in the volume becomes smaller as one gets closer to the zero spatial
volume. Consequently, in principle, one might need take an infinite
number of steps so that the spatial volume collapses to zero.
We believe that one reason behind this behaviour is the ``no collision" requirement as set by Barrett et al.
Figure (\ref{5and16-CellGraphLR}) shows the outcome of our numerical example for the 5-cell and 16-cell lattices. It is quite evident that although we have worked with much cruder underlying lattices, the evolution of
our model is in good agreement with the analytical solution.

\section{Conclusion}\label{Conclusion}
In this paper, we have shown how to account for causality in the
``Parallelisable Implicit Evolutionary Scheme''. We obtained the
relevant Regge equations that must be used in determining the unknown
edge-lengths in this evolutionary method. The evolution of a skeletonised
FLRW universe, using a 5-cell triangulation of a 3-sphere and a 16-cell
triangulation, was examined. It was shown that the results of this
approximation are in good agreement with the analytical model. The
inclusion of causality in PIES is indeed an important step in making this
evolutionary method into an excellent probe in investigating the evolution
of complicated manifolds including those with non-trivial topology.


\appendix
\setcounter{section}{1}
\section*{Appendix}
\subsection{Area of a SST Triangle}\label{Area of a Time-Like Bone}
In Euclidean geometry, the area of any triangle can be obtained using
Heron's formula. For a triangle with edges a, b and c, Heron's
formula reads:
\begin{equation}
Area = \frac{1}{4}\sqrt{P(P-a)(P-b)(P-c)}
\end{equation}
where $2P$ is the perimeter of the triangle. This formula can be
written in terms of the determinant of a matrix, known as the Cayley-Menger
determinant:
\[ Area^2 = -\frac{1}{16} \left|
\begin{array}{cccc}
0&1&1&1\\
1&0&b^2&a^2\\
1&b^2&0&c^2\\
1&a^2&c^2&0
\end{array}
\right|. \]

It is possible to show that Heron's formula holds true for triangles
on the Minkowski plane. The only catch is that instead of the length of
the edges (a positive value) one has to put the norm of the edges, in
particular account whether an edge is time-like, null or space-like. In
addition, to account for the original classification of the bones by
Regge \cite{Ref1}, the negative sign, in front of the determinant must be
eliminated.
Using the Cayley-Menger determinant, the area of (SST) Triangle,
$\triangle{ABC}$, shown in figure (\ref{Triangle4}) is given by:
\begin{equation}
\label{area of a SST bone}
A^2_{\triangle{CAB}} = \frac{1}{16}\big((a^2 + c^2 + b^2)^2 - 4c^2b^2\big)
\end{equation}


\subsection{Area of a NSS Triangle}\label{Area of a NSS Triangle }
Another type of triangle whose area is required in obtaining equation
(\ref{Area of CBP}) is a triangle with two space-like sides and one null
side (NSS). $\triangle{CAM}$ in figure (\ref{Triangle3}) is a (NSS) triangle with sides
$\wvec{CA}$ and $\wvec{AM}$ space-like and $\wvec{CM}$ null.
The area of a (NSS) triangle can be obtained in a similar manner using
a Cayley-Manger determinant:
\begin{equation}
\label{Area CAM1}
A^2_{\triangle{CAM}} = \frac{1}{16}(m + c)^2(m - c)^2
\end{equation}
For the purpose of this work however, we require to write this area in
a different form:

\begin{figure}[htbp]
\begin{center}
\includegraphics[scale=0.4]{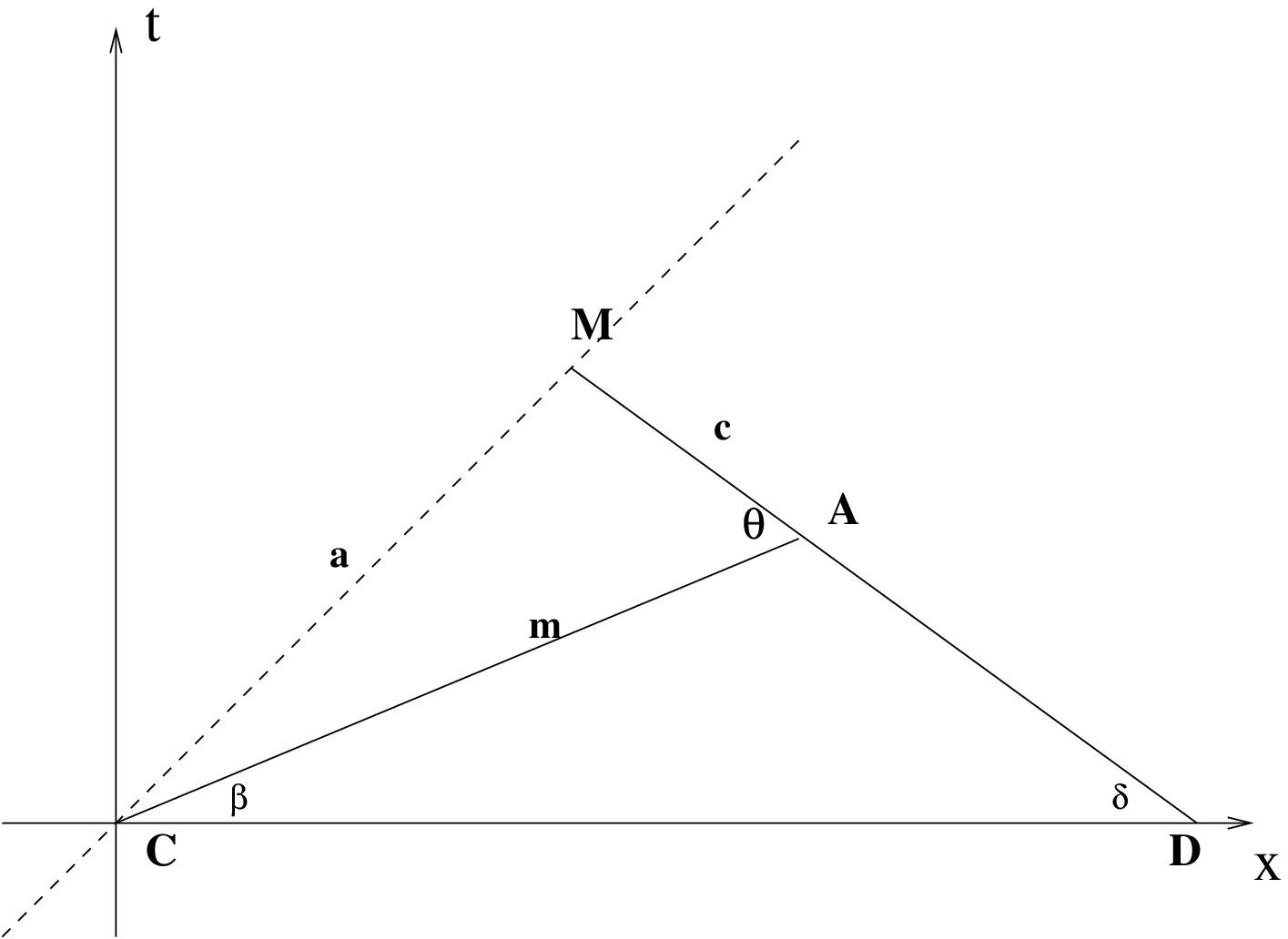}
\end{center}
\vspace*{8pt}
\caption{\label{Triangle3}
$\triangle{CAM}$ is a NSS Triangle with $\wvec{CM}$ being the
null edge.}
\end{figure}
This expression of area will be given in terms of one of the space-like edges
and the angle between the two space-like edges; start with
\begin{equation}\label{Null Dot Product}
\wvec{CA} + \wvec{AM} = \wvec{CM}
\end{equation}
Taking the dot product of both sides of (\protect \ref{Null Dot
Product}) with $\wvec{CM}$ and using the fact that
$\wvec{CM}$ is null one obtains:
\begin{displaymath}
\wvec{CA}\cdot\wvec{CM} =
-\wvec{AM}\cdot\wvec{CM}
\end{displaymath}
Since $\wvec{CM}$ is a null vector in the first quadrant, it can always
be written as:
\begin{displaymath}
\wvec{CM} = a\:(\hat{t} + \hat{x})
\end{displaymath}
where ``$a$" is a real and positive number.
Then:
\begin{eqnarray*}
a\:m\:(\cosh{\beta} - \sinh{\beta}) &=& - a\:c\:(-\cosh{\delta} -
\sinh{\delta}) \\
m\:(\cosh{\beta} - \sinh{\beta}) &=& c\:(\cosh{\delta} + \sinh{\delta})
\end{eqnarray*}
Solving for ``$c$", we have:
\begin{displaymath}
c = m\:\frac{(\cosh{\beta} - \sinh{\beta})}{(\cosh{\delta} +
\sinh{\delta})}
\end{displaymath}
which can be re-written as:
\begin{eqnarray*}
c &=& m\:(\cosh{\beta} - \sinh{\beta})(\cosh{\delta} - \sinh{\delta})\\
 &=& m\:(\cosh{(\delta + \beta)} - \sinh{(\delta + \beta)})
\end{eqnarray*}
In figure (\ref{Triangle3}), $\delta +\beta = \theta$ \cite{Ref6},
and thus:
\begin{displaymath}
m = c\:(\cosh{\theta} + \sinh{\theta})
\end{displaymath}
Using this result in equation (\protect \ref{Area CAM1}), one obtains:
\begin{equation}
\label{Area CAM2}
A_{\triangle{CAM}} = \frac{1}{2} c^{2}(\cosh{\theta} +
\sinh{\theta})\sinh{\theta}
\end{equation}

We will now use a similar strategy to obtain an expression for the area of $\triangle{CBP}$ as shown in figure (\ref{Triangle4}).
In this figure, the line segment BP is null and thus divides $\triangle{CAB}$ into an
(NSS) and an (NST) triangle. To calculate the area of $\triangle{CBP}$,
it is easiest to subtract the area of $\triangle{APB}$ from that of
$\triangle{CAB}$.
Since $\triangle{APB}$ is a (NSS) triangle, its area can be obtained
using an equation similar to (\protect \ref{Area CAM2}):
\[
A_{\triangle{ABP}} = \frac{1}{2}{c^2}(\cosh{\xi} - \sinh{\xi})\sinh{\xi}
\]
It is quite clear that $\sinh{\xi} = 2A_{\triangle{CAB}}/(b\:c)$ and $\cosh{\xi} = (b^2 + a^2 + c^2) / (b\:c)$; thus the area of $\triangle{ABP}$, in terms of the edge lengths, is
given by:
\begin{equation}\label{Area ABP}
A_{\triangle{ABP}} = \frac{A_{\triangle{CAB}}}{2b^2}\big((a^2+b^2+c^2)
- 4A_{\triangle{CAB}}\big)
\end{equation}
Since the area of $\triangle{CBP}$ is given by:
\begin{displaymath}
A_{\triangle{CBP}} = A_{\triangle{CAB}} - A_{\triangle{APB}}
\end{displaymath}
one can use equation (\protect \ref{Area ABP}) to obtain the area of
$\triangle{CBP}$ as
\begin{equation}\label{Area CBP}
A_{\triangle{CBP}} = A_{\triangle{CAB}} -
\frac{A_{\triangle{CAB}}}{2b^2}(a^2+b^2+c^2) +
\frac{4A^2_{\triangle{CAB}}}{2b^2}
\end{equation}
The particular form of this equation facilitates the calculations of
section (\protect \ref{Variation of a Time-Like Bone with respect to a Time-Like
edge}).

\end{document}